\documentclass[twocol]{ametsocV6.1}

\usepackage{url}
\usepackage{enumitem}
\usepackage{amsmath}
\usepackage[font=small,labelfont=bf]{caption}

\title{Covariance based estimates of near-boundary diapycnal upwelling in a submarine canyon}

\authors{Kurt L. Polzin\aff{a}\correspondingauthor{Kurt L. Polzin, kpolzin@whoi.edu}}

\affiliation{\aff{a}{Woods Hole Oceanographic Institution, Woods Hole, Massachusetts}}

\abstract{We present direct covariance estimates of temperature flux from a conventional taut wire mooring placed in a narrow submarine canyon on the continental slope west of Ireland.  Estimates of stratification from both moored sensors and vertical profiling instrumentation are used to facilitate the interpretation of these temperature transport estimates in terms of diathermal upwelling.  At depths of 50-125 meters above bottom, the temperature flux is up-gradient, rather than down the mean gradient, and the corresponding vertical divergence of the diathermal temperature flux implies a diapycnal velocity of approximately 1 mm s$^{-1}$, consistent with diathermal migration estimated from a dye release study.  Cospectra document that the temperature flux is related to a wave breaking process under-pinned by semi-diurnal frequencies.  This up-gradient flux also results in a highly non-local temperature variance budget.  We demonstrate that nonlinear temperature variance production at the bottom boundary is the root cause, with redistribution aloft determining the temperature flux profile.   
 }

\begin{document}

\maketitle

%
%
%
%
%
%

%








\section{Introduction}
The deep ocean is a vast reservoir of heat and carbon and the processes that impact their renewal and transformation are of fundamental importance to long term Earth system behavior \citep{melet2022role}.  The processes by which cold, dense waters upwell have been an immense source of speculation and controversy since being framed in \cite{Munk1966abyssal}.  We point the reader to a recent review of the concepts \citep{polzin2022mixing} that underpin our understanding of boundary mixing and disparate opinions that have yet to be resolved.  

\begin{figure}[htb]
\noindent\includegraphics[width=0.48\textwidth]{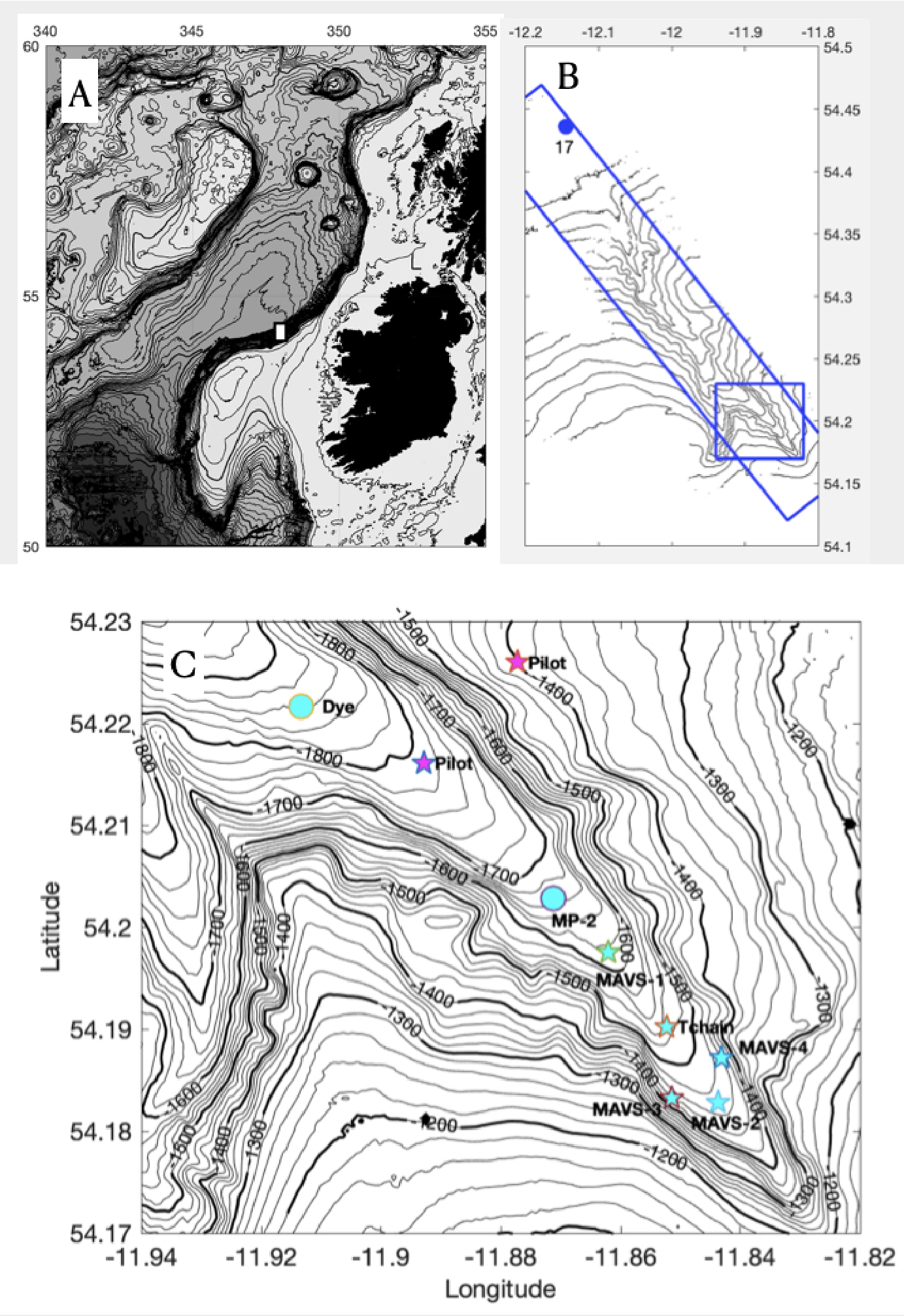}\\
\caption{ Geography of the Boundary Layer Turbulence - Recipes field program.  Panel A:  Overview of the Rockall Trough.  The moored contributions were executed in a canyon delineated by the white square just west of Ireland.  Panel B:  Overview of the canyon topography.  The blue square is expanded in panel C.  Multibeam data from the blue rectangle are used in Section \ref{sec:Discussion}.  The round symbol locates CTD 17 from the third BLT cruise.  This station is used to defined the 'interior' stratification for ray tracing purposes in Section \ref{sec:Discussion}.  Panel C:  location of the moored resources, the  dye release and 24-hour CTD/LADCP repeat stations from the UK Pilot program.  }\label{fig:1}
\end{figure}

This article presents an assessment of upwelling from direct flux estimates using moored instrumentation deployed as part of a multi-national multi-investigator field program, Boundary Layer Turbulence - Recipes, situated on the continental slope west of Ireland, figure \ref{fig:1}.  This article is one of three assessments of upwelling from the field program, one from a dye release \citep{wynne2024observations}, and a second dissipation based \citep{garabato2024convective}.  The three are linked as follows.  The dye release provides a concentration weighted assessment of the migration of the dye across temperature isopleths.  The dissipation based study is conducted by sorting estimates of temperature dissipation rate into a temperature coordinate system and quantifying the diathermal dissipation gradients by channeling a theoretical work in \cite{winters1996diascalar}.  This study is a brute force application of direct flux estimates routed through a diapycnal advection-diffusion balance, \cite{mcdougall1989dianeutral}.  

\begin{figure}[htb]
\noindent\includegraphics[width=0.48\textwidth]{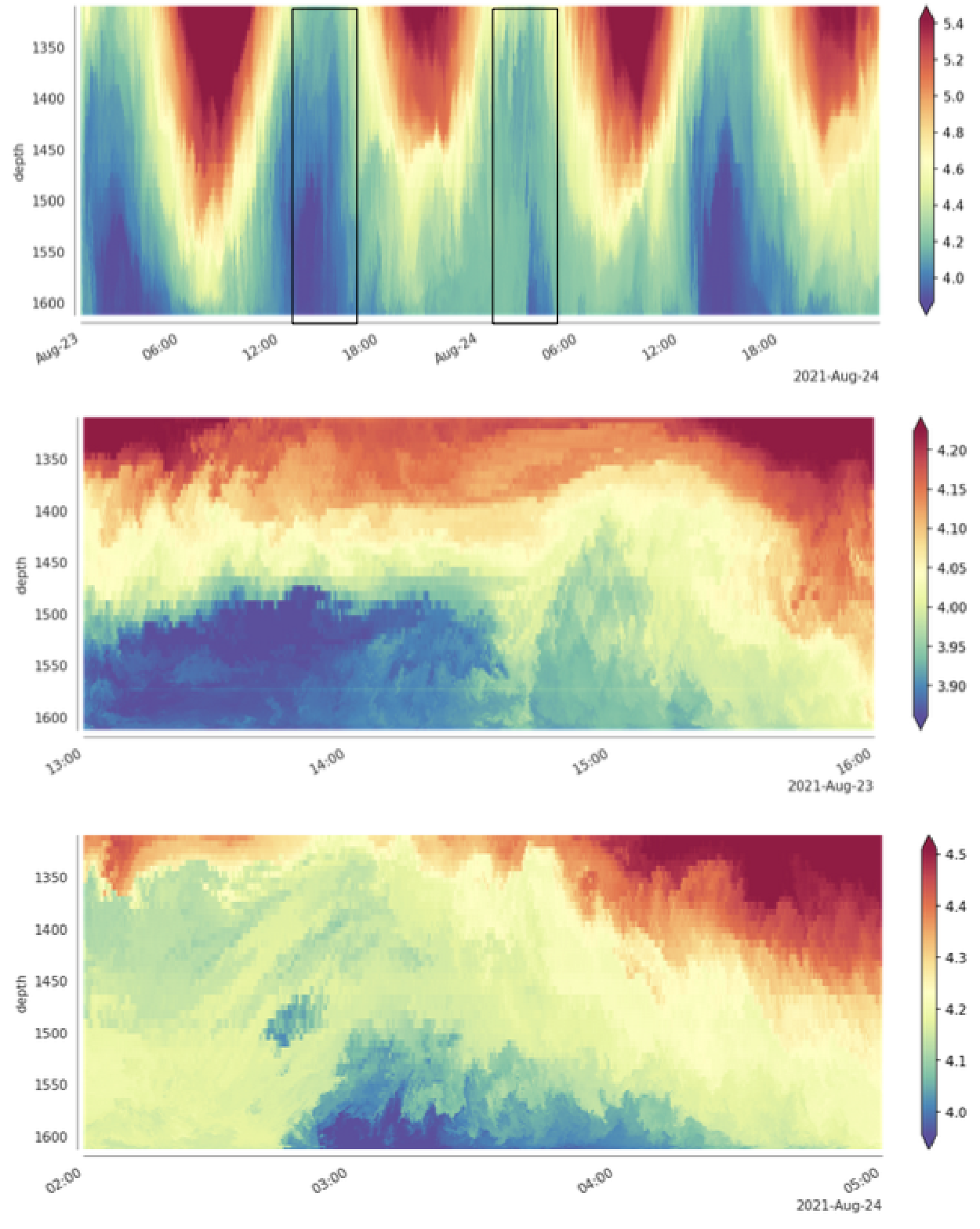}\\
\caption{Two day time series of temperature from MAVS2.  Note the dominance of semidiurnal time scales and repeated occurrence of inverted stratification as part of a wave breaking process.  Currently a MAVS1 placeholder from Gunnar.  upper panel wants to demonstrate that isotherms are hoisted from canyon bottom to the rim of the canyon head twice a day. 
 }\label{fig:2}
\end{figure}

Our parsing of direct covariance estimates starts by using a Reynolds decomposition of scalar conservation statements in Cartesian coordinates.  We use a time average to define mean fields:
\begin{equation}
    \overline{\dots} \; = \; \frac{1}{\tau} \int_{0}^{\tau} \dots \; dt \; ,
\label{eq:Reynolds}
\end{equation}
in which $t$ is time.  Using density $\rho$ as the conserved scalar,
\begin{equation}
    \partial_t \overline{\rho} + \overline{{\bf u}} \cdot \nabla \overline{\rho} +  \nabla \cdot \overline{{\bf u}^{\prime} \rho^{\prime}} \cong 0 \; ,
\label{eq:MeanReynolds}
\end{equation}
in which ${\bf u}$ is the three-dimensional velocity vector, the time derivative is indicated by a subscript and ${\bf \nabla}$ is the three-dimensional gradient operator.  A formula for diapycnal upwelling is obtained by transforming from Cartesian to isopycnal coordinates.  This implicitly assumes the hydrostatic relation and a one-to-one mapping between the mean density field and vertical coordinate.  This transformation brings considerable simplification as there is limited advection of density along surfaces of constant density:  
\begin{equation}
   \partial_t \overline{\rho} + w^{\perp} \cdot \nabla \overline{\rho} +  \nabla_{\perp} \cdot \overline{{\bf u}^{\prime} \rho^{\prime}} \cong 0
\label{eq:upwelling}
\end{equation}
in which we have used ${\perp}$ to render the velocity component and gradients normal to density surfaces.  Expressions that appear in \cite{ferrari2016turning} and \cite{marshall1999reconciling} contain vector notation representing the flux as minus a constant times the mean gradient.  Equation (\ref{eq:upwelling}) stands on its literal interpretation:  in steady state, the flux divergence drives diapycnal transport.  Hence we estimate the flux divergence to obtain the diapycnal transport rate.

A great deal of understanding concerning global ocean upwelling of cold dense water from the abyss is founded on the `one-dimensional model' originating in \cite{phillips1970flows, wunsch1970oceanic} and follows a trajectory of \cite{phillips1986experiment, thorpe1987current, garrett1990role, ferrari2016turning, mcdougall2017abyssal}.  That model has three significant assumptions:  
\begin{enumerate}[label=Assumption \arabic*:]
    \item the vertical flux divergence dominates the horizontal components, 
    \item the flux can be directly related to the mean gradient (a flux-gradient relation underpinned by mixing length arguments), and
    \item the mean fields evolve on a slow ($\tau \gg f^{-1}$) time scale.
\end{enumerate} 
See \cite{polzin2022mixing} for a detailed discussion of the development.  
The rational for these assumptions is quite clear.  Given the thermal wind relation on the slow time scale and simple eddy viscosity - eddy diffusivity closures for momentum and buoyancy, the entire structure of the boundary layer can be solved for, including the residual flow that represents diapycnal upwelling.  This boundary layer structure results from the no-normal flux boundary condition on density:
\begin{equation}
\overline{{\bf u}^{\prime}_{\perp} \rho^{\prime}}\big{|} _{H_{ab}=0} \;\;= 0 \; .
\label{eq:BoundaryCondition}
\end{equation} 
Since there is no normal flux to transport temperature through the boundary, Assumption 1 requires the vertical flux to vanish at the boundary for moderate topographic slopes.  Assumption 2 parlays this understanding to require weakened near-boundary stratification.  This, in turn, results in a mid-depth maximum on the buoyancy flux and, from (\ref{eq:upwelling}), upwelling hard up against the bottom boundary.

For the case of a depth independent diffusivity \citep{thorpe1987current}, one obtains an exponential dependence $e^{-q \hat z}$ in slope normal coordinate $\hat z$:
\begin{equation}
q^4=\frac{1}{4} \Bigg[ \frac{N^2 {\rm sin}^2(\theta)}{\nu_e \kappa_e} + \frac{f^2}{\nu_e^2} \Bigg]
\label{eq:scaleHeight}
\end{equation}
in which $\theta$ is the angle of topographic repose, $N$ stratification, $f$ Coriolis frequency, and $\nu_e$ and $\kappa_e$ are eddy viscosity and eddy diffusivities in flux-gradient closures.  

The model brings with it some thematic difficulties, though, that are encapsulated in the kinematics of the waves likely responsible for boundary mixing \citep{polzin2022mixing}.  These are situations in which ray trajectories are critical, or nearly so, so that particle trajectories nearly parallel the slope.  If the one-dimensional model were true, the bending of isopycnals into the slope to meet the no-flux bottom boundary condition implies a significant contribution of the horizontal fluxes to the divergence, i.e. Assumption 1 is inconsistent with the phenomenology of internal wave breaking \citep{polzin2022mixing}.  Second, near-critical motions are prone to overturning mean isopycnals and giving rise to non-local fluxes which are inconsistent with Assumption 2.  The breaking phenomenology is easily visualized using time series from temperature recorders densely distributed along a mooring cable, figure \ref{fig:2}.  Ray trajectories for this internal tide are nearly parallel the mid-canyon topography.  

This paper presents evidence based upon direct temperature flux measurements from conventional taut wire moorings that reveal Assumptions 1 and 2 are not true, Section \ref{sec:Results}.  Application of the diapycnal advection:diffusion balance in (\ref{eq:upwelling}), also presented in Section \ref{sec:Results}, returns an estimated upwelling rate of $O(1 $ mm s$^{-1})$.  This is consistent with the concentration weighted upwelling rate from a dye release immediately preceding the mooring deployment.  The kinematics of the internal tide motions that support the observed temperature flux patterns are explored in Section \ref{sec:Discussion}.  A Methods section, Section \ref{sec:Methods}, provides information about the observational campaign, ambient hydrographic patterns and tidal forcing, and statistical analyses.  A Perspectives section, Section \ref{sec:perspectives}, rounds out the narrative.   Major conclusions are summarized in Section \ref{sec:Conclusions}.  

\section{Methods}\label{sec:Methods}

note to self:  cite data reports.  

\subsection{Interior Stratification}

\begin{figure}[t]
\noindent\includegraphics[width=0.48\textwidth]{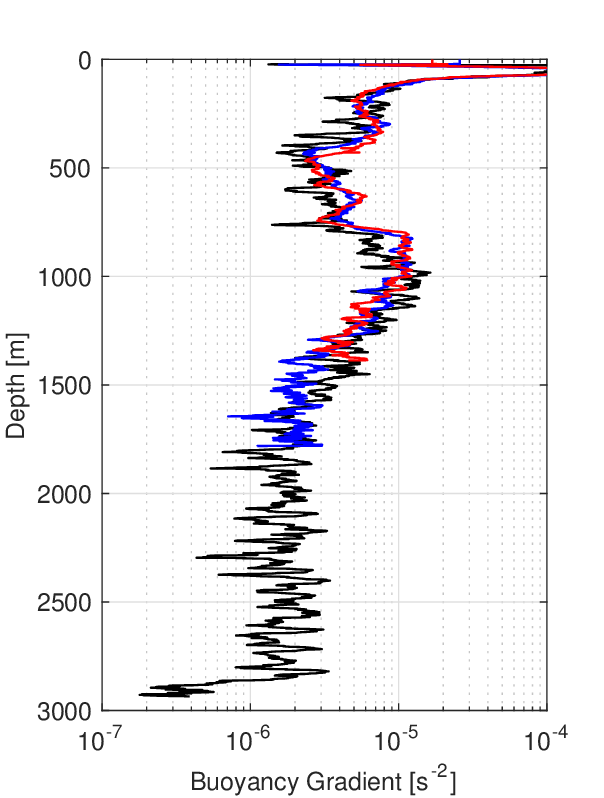}\\
\caption{20 dbar stratification, CTD 17 BLT-3 (DY153), BLT UK-Pilot Deep (DY108 stations 27-48) and UK-Pilot Shallow (DY108 stations 7-26). }\label{fig:3}
\end{figure}

We require estimates of interior (farfield) stratification to conduct ray-tracing assessments.  This is accomplished using CTD station 17 from the third BLT cruise, DY153, figure \ref{fig:1}, with 20 m first difference stratification presented in figure \ref{fig:3}.  Depicted in figure \ref{fig:3} are similar 20 m estimates obtained from 24 hour time series as part of DY108, which we refer to as the UK-Pilot study.  The traces are quite similar.  Note that the deepest profile, taken over relatively flat topography, exhibits weakened stratification at the bottom.  We are hard pressed to define a similar buoyancy anomaly at the 24 hour time series within the canyon environs, which in turn is situated on the continental slope.  This is entirely consistent with the global assessment of \cite{banyte2018weakly}.  Gazing upon the boundary layer scaling (\ref{eq:scaleHeight}), we find an ambiguous dependence upon $\theta$, with steeper slopes leading to smaller boundary layer scale heights, countered by enhanced mixing over steeper slopes \citep[e.g.][]{polzin1997spatial} leading to greater boundary layer scale heights.  To pre-view our results, an order of magnitude near-bottom estimate of $q^{-1}$ is approximately 20 m, similar to the distance of closest descent of the wire lowered profiler (10 m) and the 20 m first difference.  We will do much better with moored estimates of stratification.   

\subsection{MAVS moorings and current meters}
We focus on data from the MAVS2 mooring, in which there are 8 Modular Acoustic Velocity Sensors (MAVS) distributed at nominal heights above bottom of 4.7, 12, 26, 50, 76, 126, 200 and 275 m along a 300 m tall mooring deployed in 1466 m water depth.  There are approximately 80 temperature recorders taped onto the mooring cable and integrated into the current meters.  We also use temperature recorder data from the MAVS1 mooring.  The data return from the MAVS1 current meters is less complete than that of the MAVS2 mooring.  MAVS1 and MAVS2 were recovered in October 2021 and redeployed as MAVS3 and MAVS4.  

We employ standard MAVS5 acoustic travel time current meters produced by Nobska Development Corp. and described in a plethora of white papers available on their web site, \url{https://nobska.net}.  We upgrade the sensor by providing external power through a 120 D-cell battery pack and log serial data from an RBR temperature device hosting a custom 10 cm long sensing sting whose tip is placed within the 10 cm$^3$ sampling volume of the current meter.  We sample at 5 Hz, the maximum rate possible with the 1980's CF2 based computer architecture.  The Alkaline battery packs provide nominal 72-75 days records.  

The character of these data are described below.  

\subsubsection{Spectra}
Record averaged spectra reveal high frequency velocity spectra consistent with a $-5/3$ inertial subrange and approximate 1/2 cm s$^{-1}$ white noise.  Anomalous peaks at high frequency are likely associated with vortex induced motions.  Close inspection suggests the vertical velocity component is lower within the inertial subrange than the horizontal.  On shore studies conducted in a turbulent plume document that the longitudinal velocity component reports low by 10-20\%.  This velocity component projects onto the vertical during normal deployment.  We do not correct for this bias in this paper.  

\begin{figure}[t]
\noindent\includegraphics[width=0.48\textwidth]{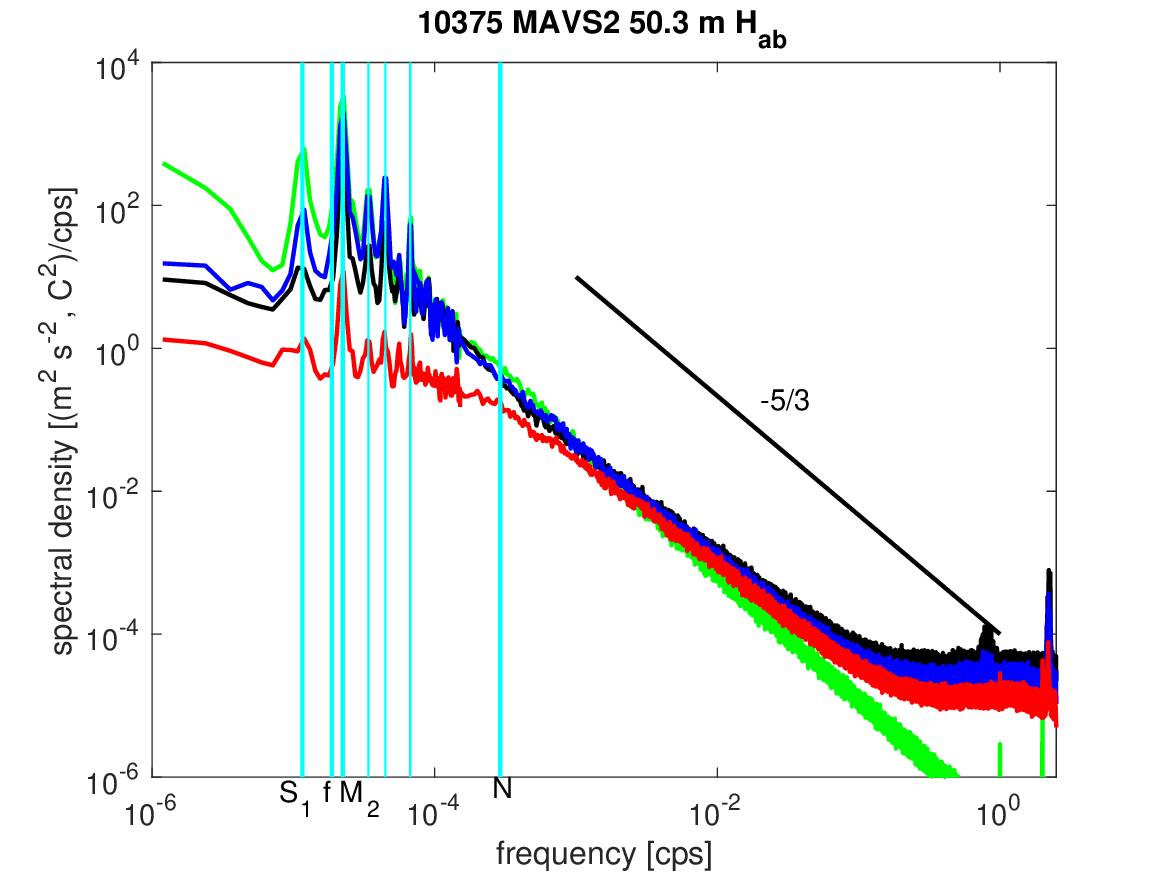}\\
\caption{ 10375 spectra \@ 50 m $H_{ab}$.  Black (east), blue (north), red (up) and green (temperature normalized by $N/\theta_z$ so as to be proportional to potential energy).  Note that the red trace tends to lie below the black and blue traces by (10-20\%)$^2$. This is understood to be a bias for currents parallel to the longitudinal axis of the sensor.  Solar diurnal, $S_1$; Coriolis, $f_0$; lunar semi-diurnal, $M_2$; and buoyancy frequency $N$ are delineated as thick vertical lines.  The thin vertical lines delineate harmonics: $S_1+f_0$, $2M_2$ and $3M_2$.   }\label{fig:4}
\end{figure}

\subsubsection{Temperature-Velocity Cospectra}
\begin{figure}[t]
\noindent\includegraphics[width=0.48\textwidth]{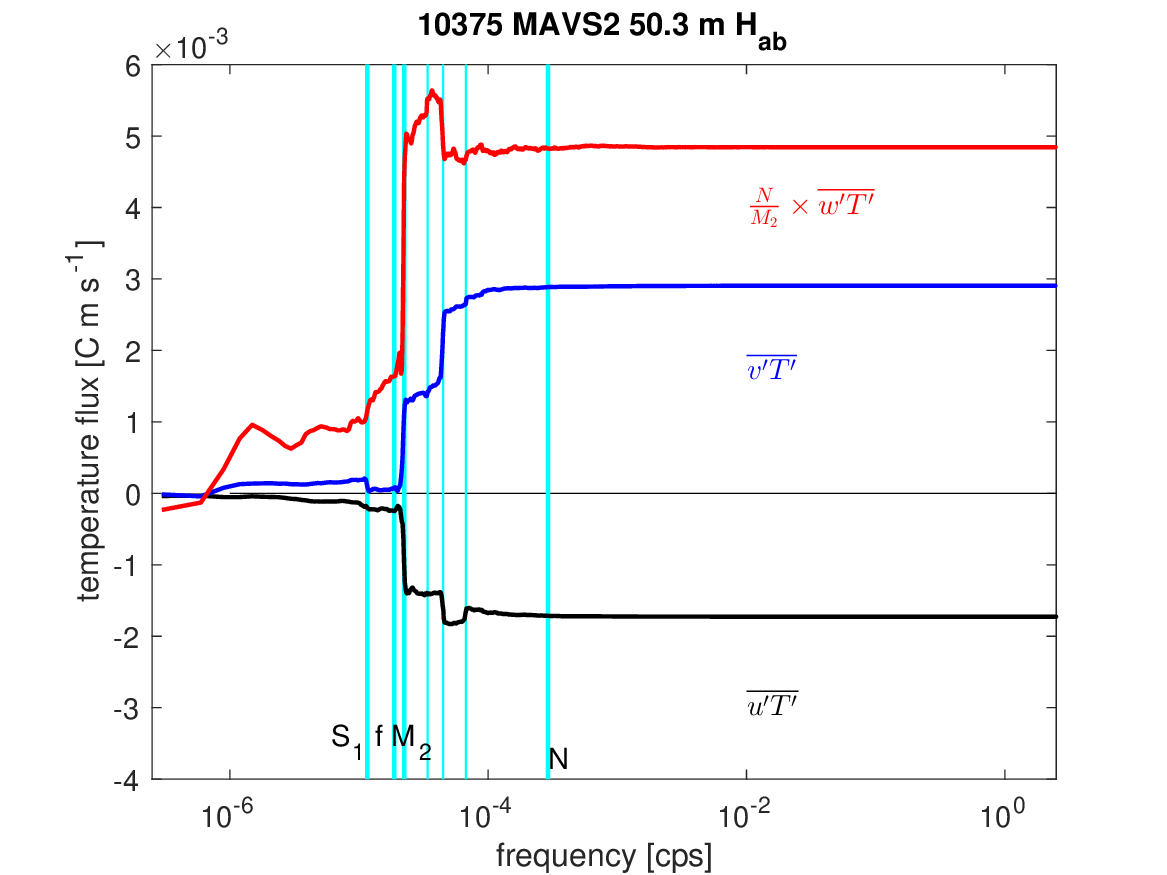}\\
\caption{ 10375 temperature cospectra. Black (east), blue (north), red (up).  Solar diurnal, $S_1$; Coriolis, $f_0$; lunar semi-diurnal, $M_2$; and buoyancy frequency $N$ are delineated as thick vertical lines.  The thin vertical lines delineate harmonics: $S_1+f_0$, $2M_2$ and $3M_2$. }\label{fig:5}
\end{figure}

Three dimensional temperature flux spectra are presented in figure \ref{fig:5}.  We consider these as fully resolved in that contributions at low frequencies and high frequencies are small.  The bulk of the covariance accumulates at semidiurnal and several super harmonics.  Some 20\% of the vertical covariance accumulates in a frequency band between the inverse of a fortnight and $f$.  Contributions from super-buoyancy frequencies are small.  Thus, turbulent production here is underpinned by an internal tide breaking process.  

Note that the vertical flux in this record is positive.  Since potential temperature increases upward, this means that the observed vertical flux is up-gradient and decidedly inconsistent with normal conceptions of a flux-gradient closure, i.e. a diffusive process.  

\subsubsection{Convergence in time}
\begin{figure}[t]
\noindent\includegraphics[width=0.48\textwidth]{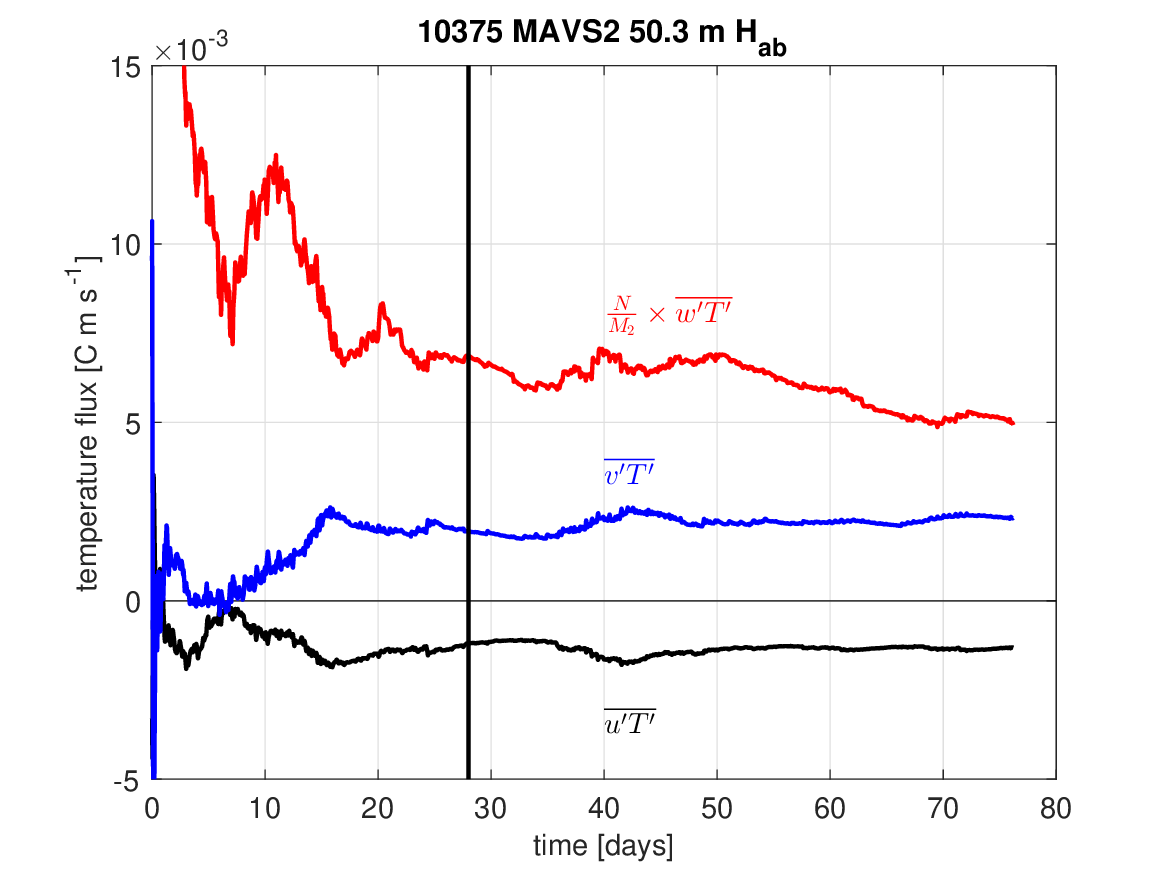}\\
\caption{ 10375 temperature covariance time series.  The covariance estimates converge in about a fortnight.  The analysis time period is broken into two time periods.  }\label{fig:6}
\end{figure}

Running integrals of temperature covariance demonstrate that it takes about a fortnight for flux estimates to converge, figure \ref{fig:6}.    In this presentation we have demeaned both temperature and velocity time series using record length estimates (\ref{eq:Reynolds}).  The results are consistent with what one might naively anticipate from the cospectra in figure \ref{fig:5}:  Barring long-term nonstationary conditions, statistical convergence is largely a matter of building up sufficient degrees of freedom at the tidal frequencies to provide a robust estimate of the temperature-velocity coherence there.  

What we have begun to appreciate, however, is that not all fortnights are created equal at every current meter, i.e. there are nonstationary tendencies in some of the records.  We tentatively relate this to amplitude dependent changes in the vertical structure of the wave-breaking process and leave that for further detailed investigations but provide hints in later sections.  We tentatively identify two distinct periods involving the first two fortnights and the last three and present those separately.  Our focus is on the first since these data are most nearly contemporaneous with the dye release.    

\subsection{Tidal Forcing and Response}

\begin{figure}[t]
\noindent\includegraphics[width=0.48\textwidth]{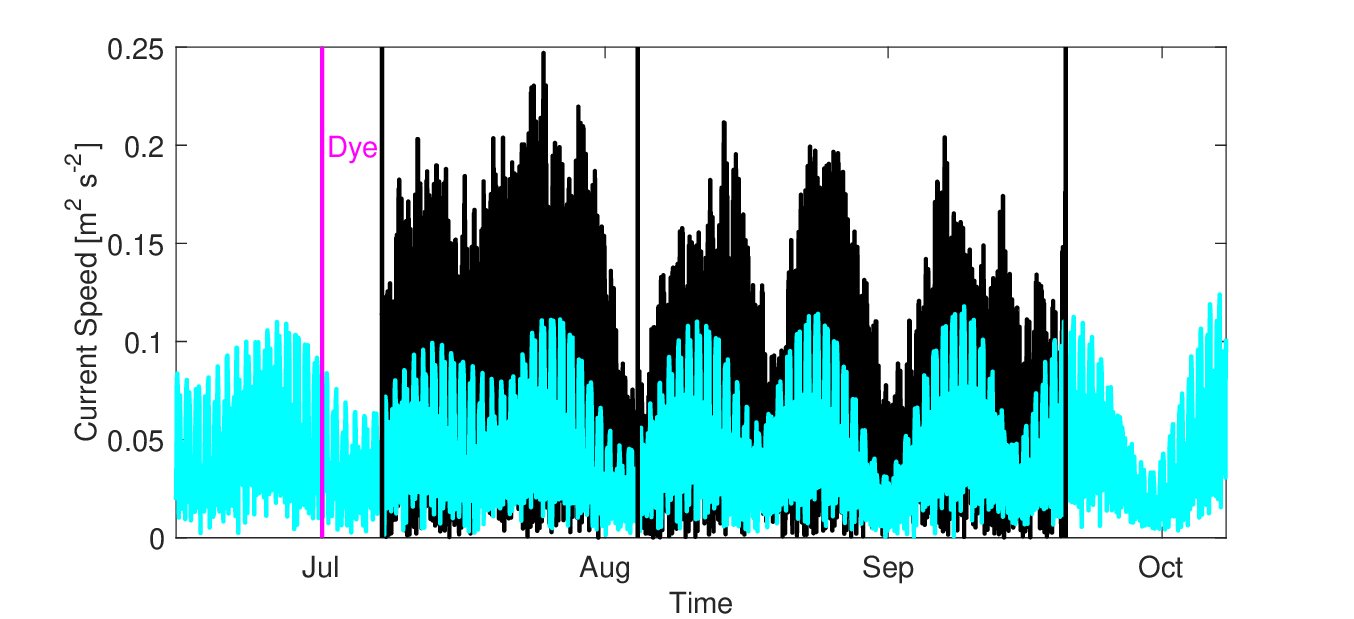}\\
\caption{ TPXOv9.5 tides (cyan) and semi-diurnal band-passed (black) currents from current meter 10289 at 200 m $H_{ab}$.  }\label{fig:7}
\end{figure}

The most cursory inspection of the data, e.g. figure \ref{fig:2}, reveals the presence of a significant semi-diurnal response.  This internal response is 'large' relative to the barotropic tide obtained from the TPXO model, figure \ref{fig:7}.  Here we have used all 15 constituents: 2n2 k1  k2  m2  m4  mf  mm  mn4 ms4 n2  o1  p1  q1  s1  s2, and note a significant diurnal surface tide that complicates the fortnightly modulation associated with the semi-diurnals.  The observed semi-diurnal response mimics the envelope of the surface tide and slightly lags in time.  The amplitude of the observed semi-diurnal tide is far larger than the full barotropic, but appears to retain the character of the full surface tide, at least in the first two fortnights of the record.  We will use these first two fortnights to investigate the temperature - velocity covariance in section \ref{sec:Results} and offer some thoughts about the amplitude of the internal response in section \ref{sec:Discussion}.  

\section{Results}\label{sec:Results}
\subsection{Stratification}
\begin{figure}[t]
\noindent\includegraphics[width=0.48\textwidth]{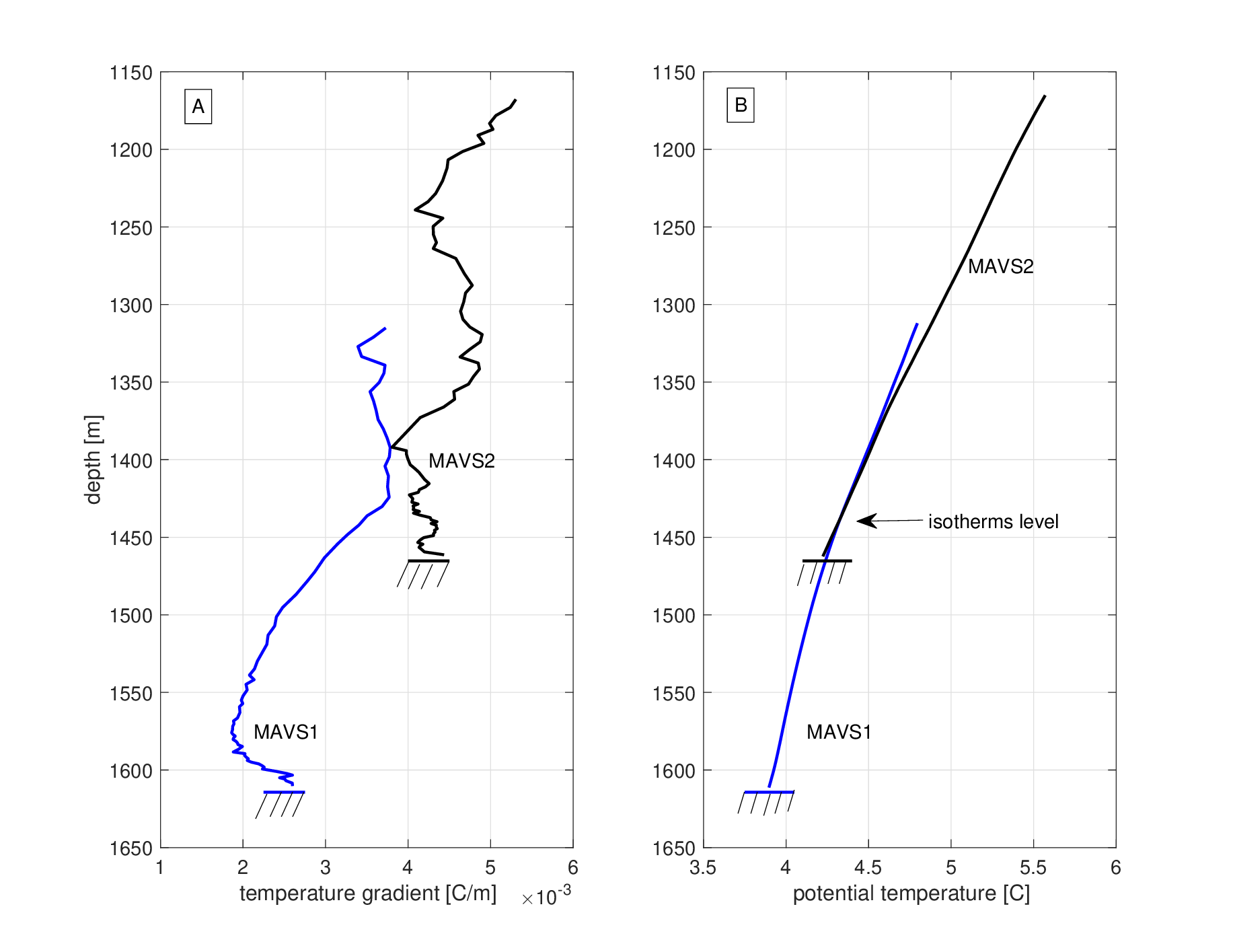}\\
\caption{ Moored stratification metrics from MAVS1 and MAVS2.  Panel A provides mean gradients for the first two fortnights.  Mean potential temperature profiles over thhis same period are rendered in panel B.  The two temperature profiles intersect at 28 m $H_{ab}$, implying that the isotherms are level at the height.   }\label{fig:8}
\end{figure}

We present vertical profiles of time mean potential temperature and potential temperature gradients from the two MAVS moorings, figure \ref{fig:8}.  Each 300 m tall mooring has approximately 80 temperature recorders with the bottom most located on the acoustic releases, about 3 m $H_{ab}$.  

Comparison of the MAVS1 and MAVS2 temperature profiles reveals the time mean near bottom stratification at MAVS2 is comparable to the interior stratification at the deeper mooring, MAVS1, figure \ref{fig:8}a.  In particular, there appears to be little near-boundary buoyancy deficit associated with the no-flux bottom boundary condition.  This is also true for the Tchain mooring, \cite{van2024near}

Plotting the two time mean potential temperature profiles, figure \ref{fig:8}b, documents an intersection at 28 m $H_{ab}$ of the MAVS2 mooring.  Thus, at this depth, the isotherms are level.  Departures above and below this level are small in the sense that the inferred isopycnal slopes are an order of magnitude smaller than the topographic slope (about 1/14).  

The fundamental prediction from the `one-dimensional model', that of reduced near-bottom stratification in association with isotherms dipping into the boundary, is not supported by this data set.  

\subsection{The Bottom Boundary Condition}

\begin{figure}[t]
\noindent\includegraphics[width=0.48\textwidth]{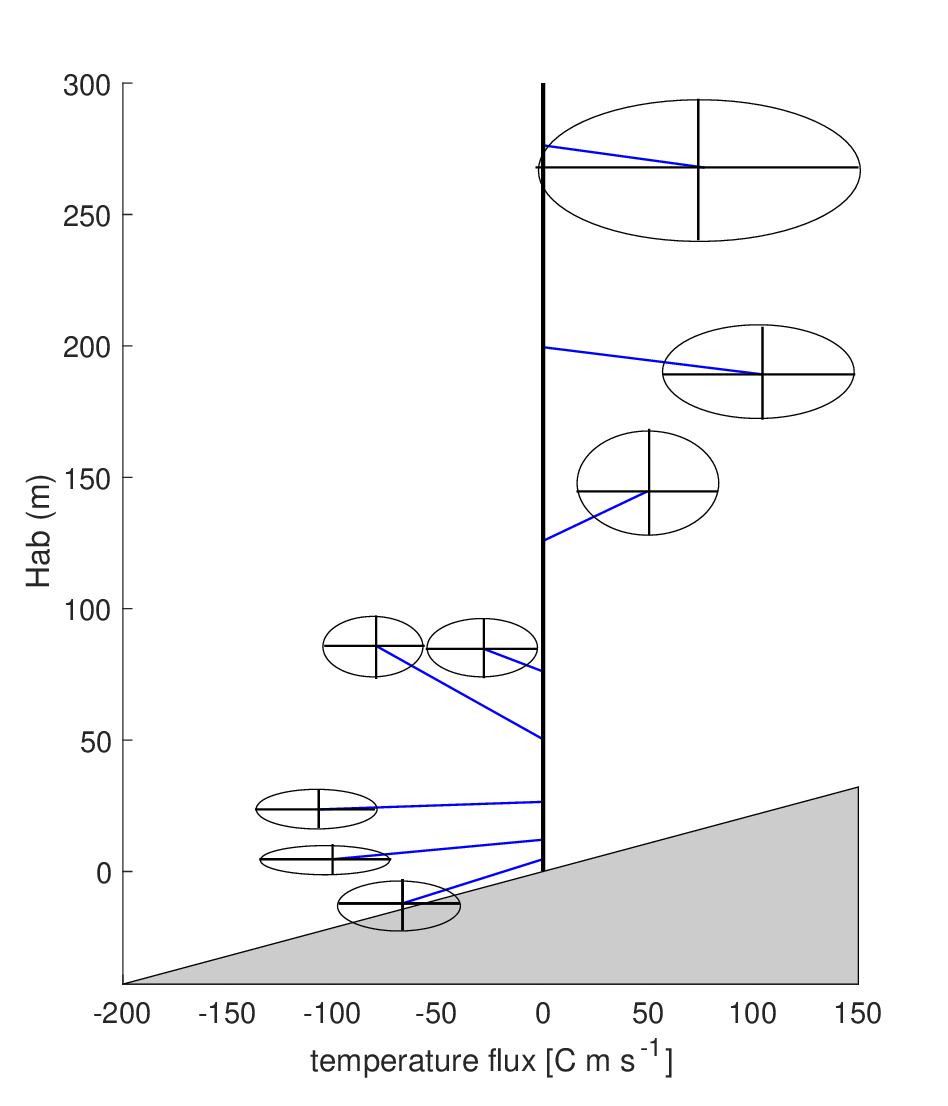}\\
\caption{MAVS2 record length along-canyon / vertical temperature flux vectors.  Ellipses denote 95\% bootstrapped uncertainty metrics.  Panel A represents the first 2 fortnights, panel B the last 3. }\label{fig:9}
\end{figure}

In this subsection we present a vertical profile of along-canyon / vertical temperature flux vectors, figure \ref{fig:9}.  As the bottom is approached, these flux vectors point down canyon and nearly parallel the slope. This satisfies the no-flux bottom boundary condition (\ref{eq:BoundaryCondition}).  In particular, while the flux magnitude decreases slightly as the bottom is approached, there is no indication that the vertical component of the flux vanishes at $H_{ab}=0$.  

\subsection{Vertical Structure of the Vertical Flux}

The preceding subsections have demonstrated that the temperature fluxes are near-parallel to the topograhic slope and that isoycnal slopes are small relative to the topographic slope.  

\begin{quote} Thus, the horizontal fluxes $\overline{u^{\prime}T^{\prime}}, \overline{v^{\prime}T^{\prime}}$ project weakly onto the diapycnal normal and the dipaycnal divergence is dominated by the vertical component, $\overline{w^{\prime}T^{\prime}}$.  Diapycnal upwelling (\ref{eq:upwelling}) is thus in response to the vertical gradient of the vertical fluxes (\ref{eq:DiapycnalVelocity}). \end{quote}

\begin{figure}[t]
\noindent\includegraphics[width=0.48\textwidth]{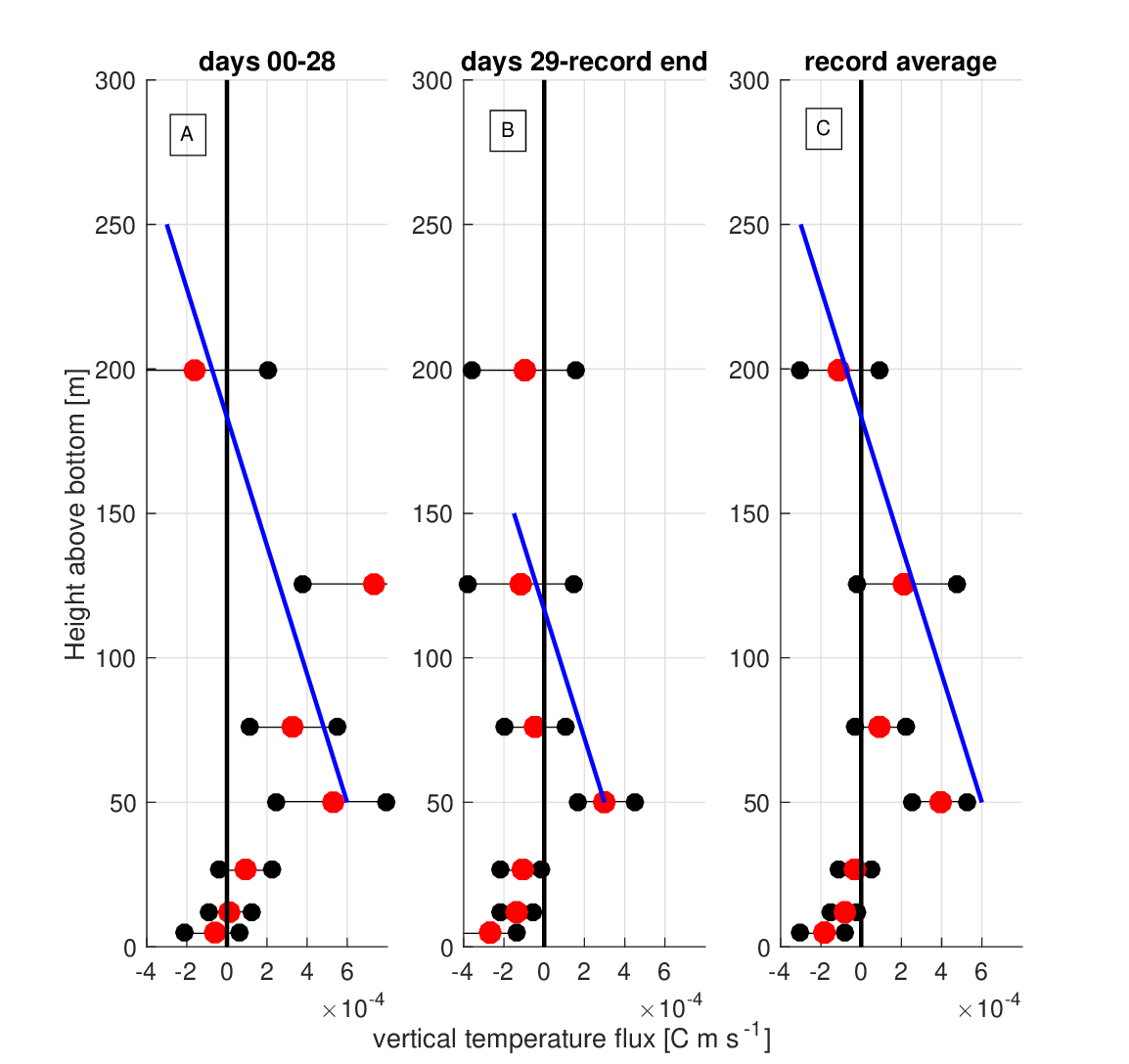}\\
\caption{Vertical profile of MAVS2 vertical flux estimates.  Panel A represents the first 2 fortnights, panel B the last 3.  Red symbols are means and black 95\% confidence intervals based upon a bootstrap algorithm.  Blue lines represent a 1 mm s$^{-1}$ diapycnal upwelling rate. }\label{fig:10}
\end{figure}

The vertical profile of vertical temperature flux is presented in figure \ref{fig:10}.  The vertical flux is positive at mid-depth, 50-125 m $H_{ab}$, negative above and below, with an increasing negative tendency into the bottom.  Both the positive flux estimates and non-vanishing vertical flux as the bottom are approached are inconsistent with the 'one dimensional model'.  

Putting a ruler onto the dots from 50 to 125 m Hab, we infer an upwelling rate of $O(1$ mm s$^{-1})$.  This is similar to the $O(1 $ mm s$^{-1})$ estimated from the preceding dye release study \citep{wynne2024observations}.  A similar technique implies downwelling within the bottom most 50 m.  Note the following:  First, in constructing a volumetric watermass or tracer budget, the effective areas and volumes below 50 m $H_{ab}$ are far smaller than those above.  Second, tracer dispersion is a Lagrangian concept and water parcels have very large vertical excursions on a wave period, figure \ref{fig:2}, relative to the vertical structure of the Eulerian statistics, figure \ref{fig:10}.  Without having more complete information about the spatial structure of the temperature fluxes, a direct comparison is naive.  

\begin{eqnarray}
    \partial_t\overline{\theta} + w^{\ast} \overline{\theta}_z & \cong & -\partial_z \overline{w^{\prime} T^{\prime}} \nonumber \\
    w^{\ast} & \cong & -\partial_z \overline{w^{\prime} T^{\prime}} / \overline{\theta}_z 
\label{eq:DiapycnalVelocity}
\end{eqnarray}

\section{Discussion}\label{sec:Discussion}

\begin{figure}[t]
\noindent\includegraphics[width=0.48\textwidth]{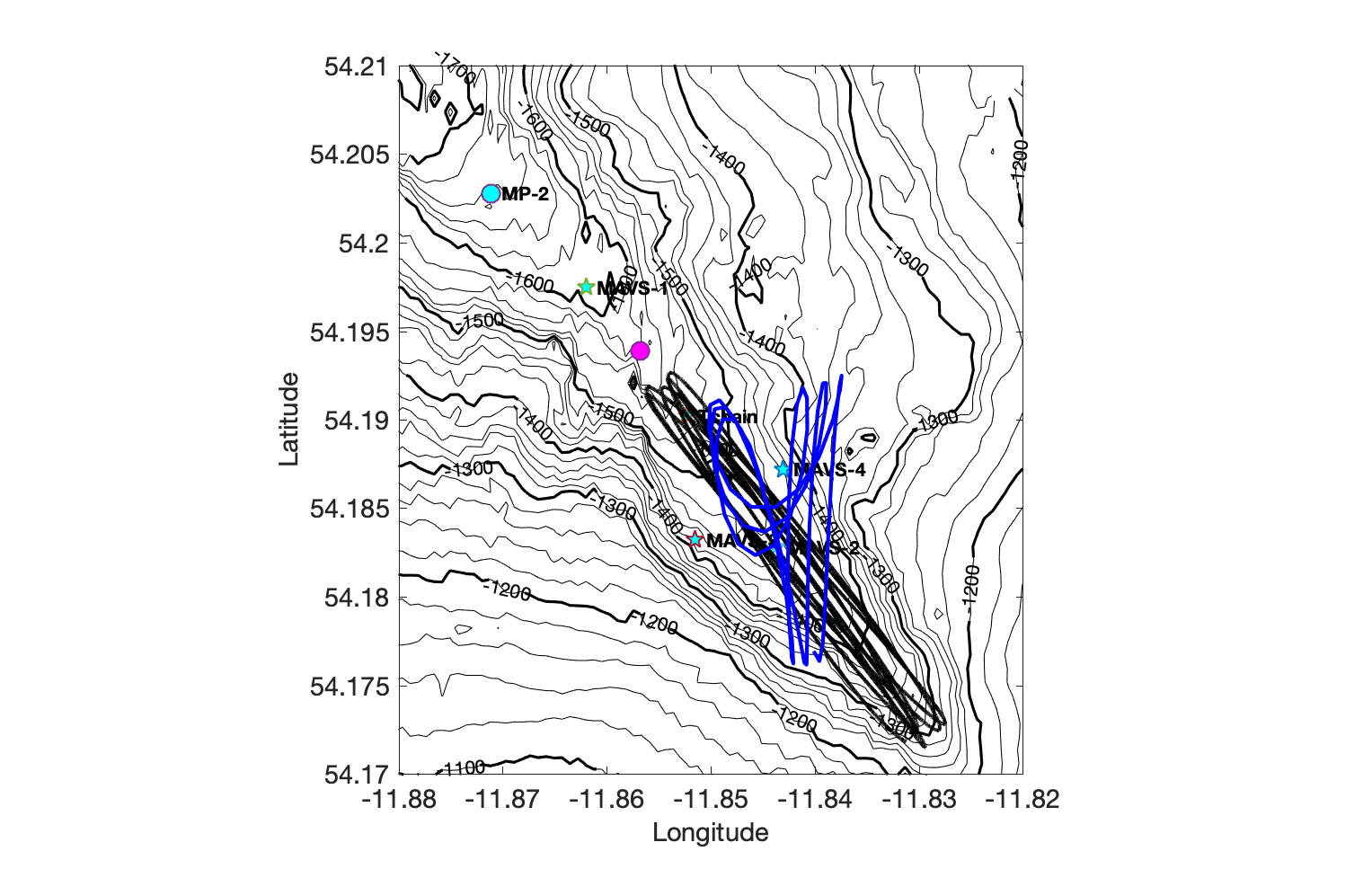}\\
\caption{Barotropic tides (blue traces) and band-passed semidiurnal horizontal currents at MAVS2 (black).  }\label{fig:11}
\end{figure}

Various opinions have been articulated about this boundary mixing problem that invoke drag (e.g. Lorke, Ruan) and differential advection (e.g. Alberto/Alford) in the context of parallel shear flows.  The parallel shear flow paradigm appears germane as currents within the canyon are nearly rectilinearly polarized, figure \ref{fig:11}.  Less emphasized lately has been the key role near-critical conditions play in setting the scene at the bottom boundary.  In any linear assessment of wave reflection from a planar sloping boundary, near-critical conditions imply a highly nonlinear high vertical wavenumber response, \citep[e.g.][]{thorpe1992thermal, slinn1996turbulent} that translates into bore-like features propagating at the phase speed of the large scale wave, \cite[e.g.][]{van2006nonlinear, winters2015tidally}.  The situation in the upper canyon differs in that a standing wave pattern without clear phase propagation sets up.  However, there is a periodic reduction in near-boundary stratification that results in super-critical Richardson numbers


We present a 4-step process as a hypothetical scenario:
\begin{enumerate}[label=(Step \arabic*)]
\item The set-up as regards near-critical conditions
\item Sourcing of instabilities resulting from anomalous near-boundary shear and stratification and/or sourcing of  vorticity at the bottom boundary,
\item wave radiation or advection of instabilities and vorticity off the boundary through flow separation, and  
\item evolution of instabilities/radiating waves and/or vorticity with the time dependent vertical and horizontal velocity gradients.  
\end{enumerate}

We walk the reader through these steps within this section.  



\subsection{Step 1a:  Canyon Geometry and near-criticality}

\begin{figure}[t]
\noindent\includegraphics[width=0.48\textwidth]{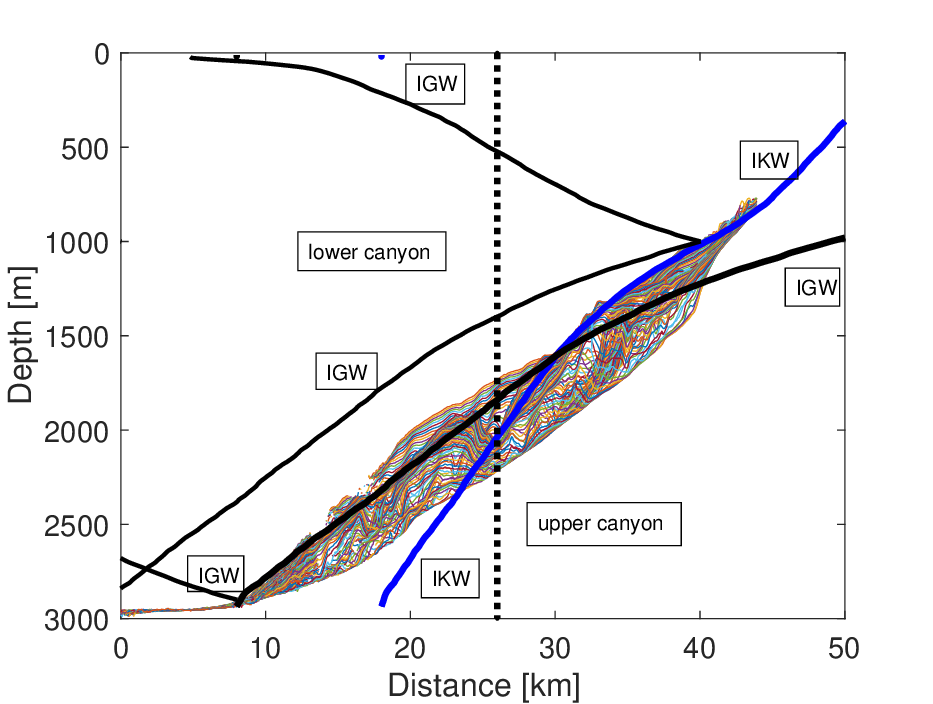}\\
\caption{thin lines represent the topography running length wise along the rectangle in figure \ref{fig:1}.  Thick black traces are internal gravity wave ray trajectories, thick black lines trajectories based upon an edge (Kelvin) wave dispersion relation.  The canyon is divided into upper and lower segments, with the upper segment being alternately sub-and super critical relative to the edge wave characteristics.  See also Yuchen.  }\label{fig:13}
\end{figure}

The canyon is incised into the continental slope such that the canyon thalweg is slightly steeper than the ray trajectories of linear internal waves with far steeper sidewalls, figure \ref{fig:13}.  In terms of a linear internal tide generation problem, one would need to couple two seaward going internal gravity waves (IGWs) to produce the elliptically polarized motions in the canyon.  While possible, as a linear problem, the magnitude of the seaward radiating baroclinic motions would not exceed the barotropic.  But the opposite inference is supported by comparing bandpassed mooring data with the TPXO estimate of barotropic motions, figures \ref{fig:2} and \ref{fig:11}.  One might relate this ratio to flow acceleration down the steeply sloping canyon sidewalls, as seen in 2-D numerical simulations \citep{sarkar2017topographic}.  However, a simpler interpretation of internal Kelvin / edge waves (IKWs) is possible.

One distinction between internal gravity waves and internal Kelvin / edge waves is that internal gravity waves have a dispersion relation of 
$$
\frac{\omega^2-f^2}{N^2} = \frac{k_h^2}{m^2}
$$
and freely propagate in 3 dimensions, whereas edge waves \citep{rhines1970edge} have a dispersion relation of 
$$
\frac{\omega^2}{N^2} = \frac{k_h^2 \; {\rm sin}^2 \phi}{m^2}
$$
where $\phi$ ($\phi = 0$ here) is the angle of the slope parallel wavenumber relative to the direction of steepest ascent and are trapped by rotation to a sloping boundary.  In this instance the sloping boundary is interpreted as the canyon sides.  An IKW solution requires the superposition of two such waves, one propagating up-slope and trapped to the right-hand wall facing up-canyon, and a second propagating down-slope and trapped to the right-hand wall facing down-canyon, \cite{ma2025standing}.  

A second distinction between IGWs and IKWs is that in the internal wave reflection problem, bore-like features appear in association with very high wavenumber content required by the boundary conditions \citep{thorpe1992thermal} traveling with a specific phase of the large scale wave, \cite{winters2015tidally}.  The proposed \citep{ma2025standing} edge solution represents a standing wave, without the phase propagation in the internal wave problem.  In this situation, we hypothesize that convergence along the boundary coupled with with the Kelvin wave seiching will serve to communicate boundary generated vorticity into the nominal interior.  We regard the potential for flow separation having either a dynamical origin or from topographic irregularities along the canyon thalweg or on the canyon walls as being high.  

The canyon itself has the appearance of a bathtub.  The fact that the IKW characteristics can be contained wholly within the upper part of the canyon provides an opportunity for wave amplitude buildup and IKW seiching that is explored in \cite{ma2025standing}
  
\begin{figure}[t]
\noindent\includegraphics[width=0.48\textwidth]{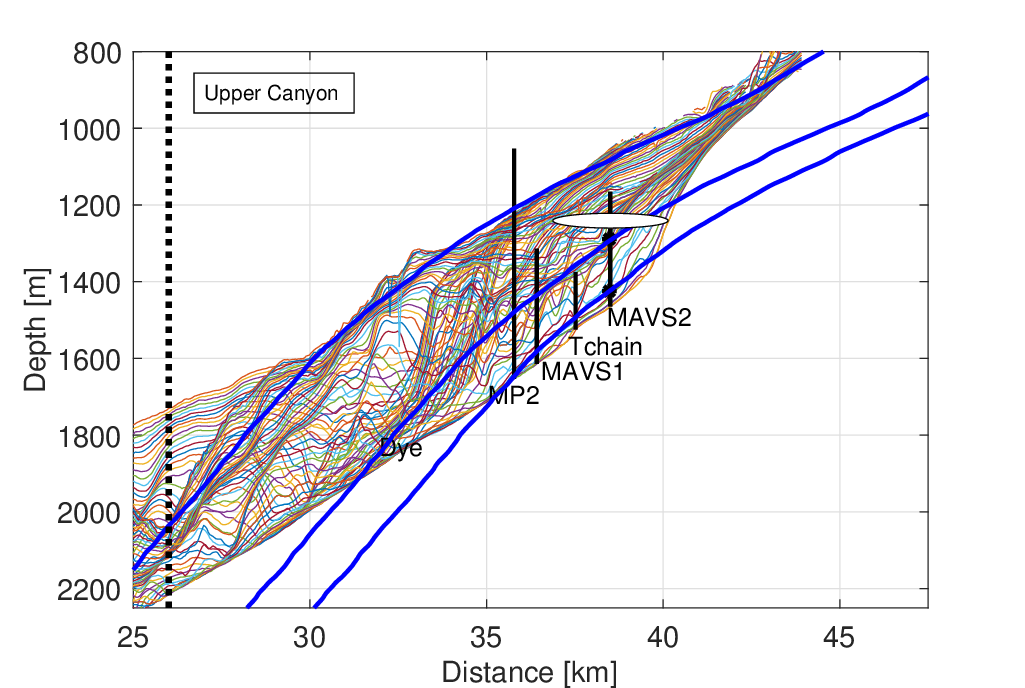}\\
\caption{As in figure \ref{fig:13}, with moored resources and dye release station located.  The white ellipse represents a characteristic internal tide displacement.  The lower rays encompass the vertical aperture of MAVS2 where upgradient temperature fluxes are noted.    }\label{fig:14}
\end{figure}

\subsection{Step 1b:  Wave Amplification}

To this point the perspective of the canyon is quite two-dimensional.  The tidal flows at the moorings are highly elliptically polarized, figure \ref{fig:11}.  In a similar fashion, the mooring data itself provides wonderful vertical and temporal resolution, but only at one location.  

We now step outside this box, and consider how observations at MAVS2 are connected to the boundaries.  One piece, based upon linear wave theory, is to acknowledge that information is carried along ray characteristics.  The edge wave characteristics intersect both the up canyon headwall and the bottom boundary down canyon from the moored assets.  Such rays provide information pathways based upon infinitesimal amplitude theory.  

\begin{figure}[t]
\noindent\includegraphics[width=0.48\textwidth]{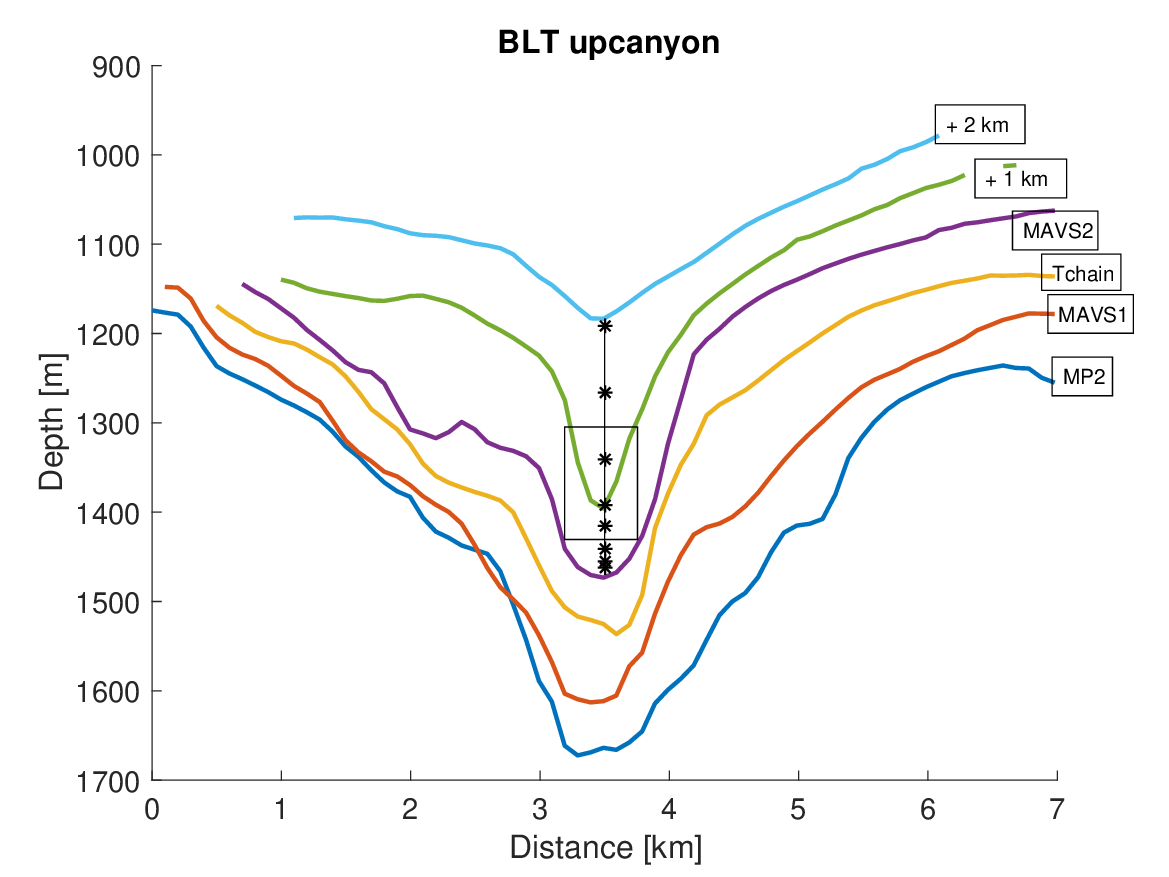}\\
\caption{An up-canyon perspective of the MAVS2 mooring.  Bathymetric profiles are taken at the MP2, MAVS1 Tchain and MAVS2 moorings, then 1 and 2 kms up canyon from MAVS2.  Note the canyon narrows at its head.  See figures \ref{fig:2} and \ref{fig:14} for information about horizontal and vertical displacements.  The square box delineates the portion of the mooring in which upgradient temperature fluxes are noted.  }\label{fig:15}
\end{figure}

A second piece are the wave amplitudes.  Peak-to-peak vertical displacements are as large as the mooring itself, figure \ref{fig:2}.  Peak-to-peak horizontal displacements are more than twice the horizontal distance to the canyon headwall, figures \ref{fig:14} and \ref{fig:15}.  At this juncture we intuit that there is significant sloping of water over the rim of bathtub shaped feature.  Associated with this sloshing of water will be large horizontal divergence and an episodic resupply of water recently in contact with the boundary.  From our understanding of boundary layer dynamics in step (i) above, that water will be loaded with spanwise vorticity.  Our understanding of vorticity dynamics suggests that interaction of that spanwise vorticity with horizontal divergence will, through a right-hand-rule, result in overturning.  

Comparison with the vertical structure of the vertical fluxes on the MAVS1 mooring could be quite instructive.  However, key current meters suffered leakage into the pressure case.  

\subsection{Step 2: instabilities}

Even though weakened near-boundary stratification is not apparent in the time mean profiles, figure \ref{fig:8}, inspection of the temperature-depth time series in figure \ref{fig:2} documents $O(1)$ modulation at a semi-diurnal period.  This sets up a situation in which the near boundary shear, likely enhanced in association with bottom drag penetration, acts in concert with near-boundary stratification reductions that exceed those aloft.  Our hypothesis is that the greater near-boundary stratification anomalies are part and parcel of the near-critical slope conditions. Together, we find low Richardson numbers result from periodic minimal stratification that occurs in conjunction with tidal shear in the near boundary region.  Such conditions are absent higher in the water column.   

\begin{figure}[t]
\noindent\includegraphics[width=0.48\textwidth]{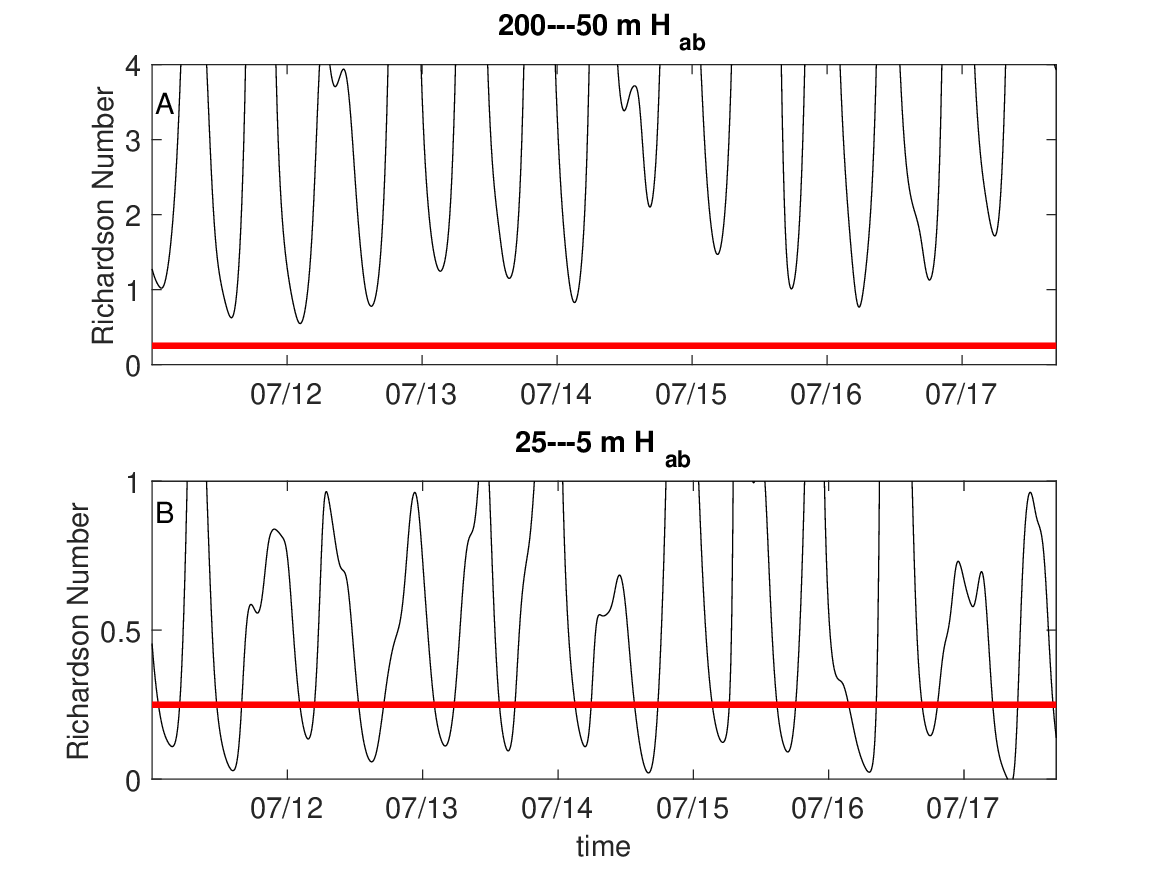}\\
\caption{ Six hour period low passed estimates of Richardson Number $R_i=N^2/S^2$ using MAVS current meters at:  Panel A:  50 and 200 m $H_{ab}$ and Panel B:  5 an 25 m $H_{ab}$.  Note the differences in vertical axis.  The red horizontal line represents the condition for linear shear instability, ${R_i = 0.25}$ }\label{fig:12}
\end{figure}

\subsection{Step 3:  Off boundary transport:  Nonlocal nonlinear temperature transport terms}

\begin{figure}[t]
\noindent\includegraphics[width=0.48\textwidth]{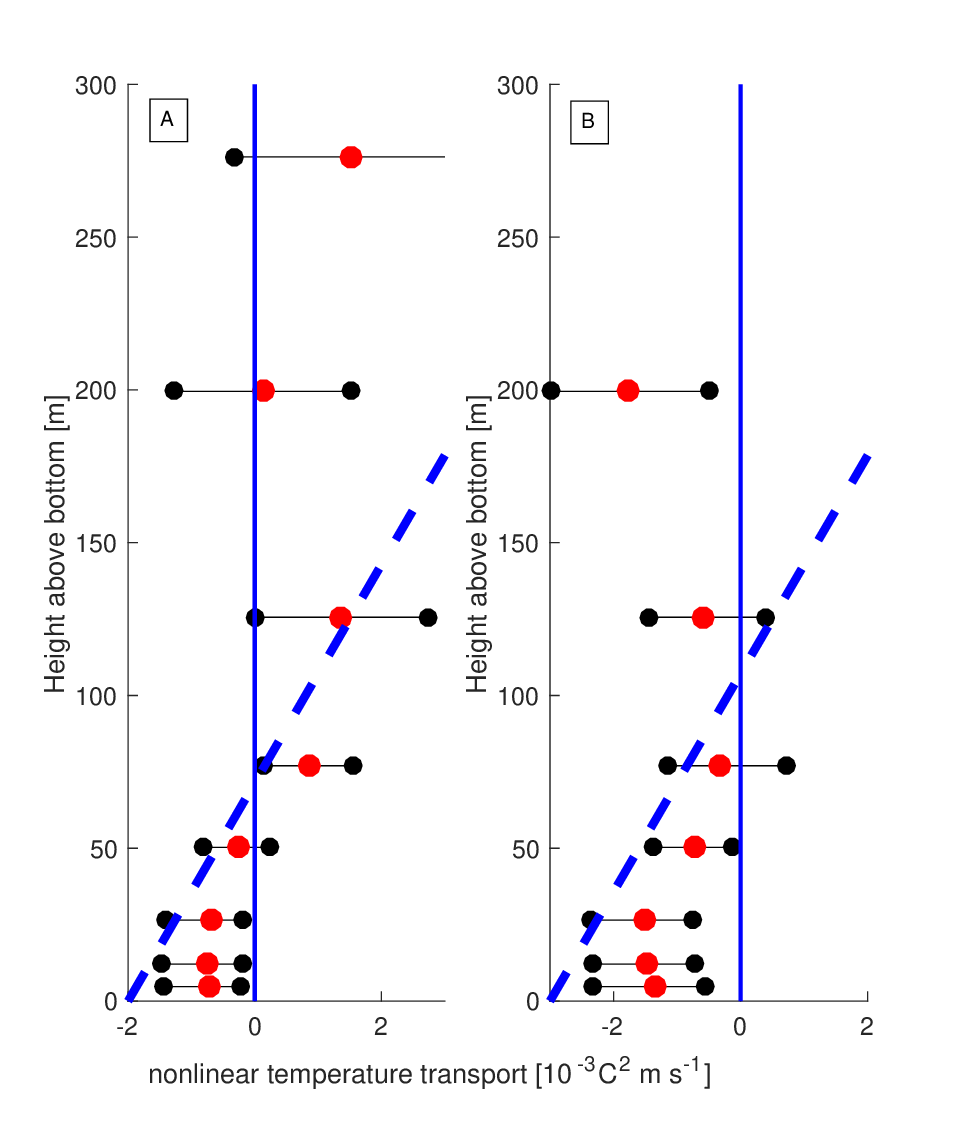}\\
\caption{ Triple correlations of temperature transport, $\overline{u_{\parallel}\theta^{\prime 2}}$, associated with along-canyon motions.  Red symbols are means and black 95\% confidence intervals based upon a bootstrap algorithm.  Panel A represents the first 2 fortnights, panel B the last 3.  The blue dashed line presents a gradient of $2\times10^{-6}$ C$^2$ s$^{-1}$ in height above coordinate.  }\label{fig:17}
\end{figure}

\begin{figure}[t]
\noindent\includegraphics[width=0.48\textwidth]{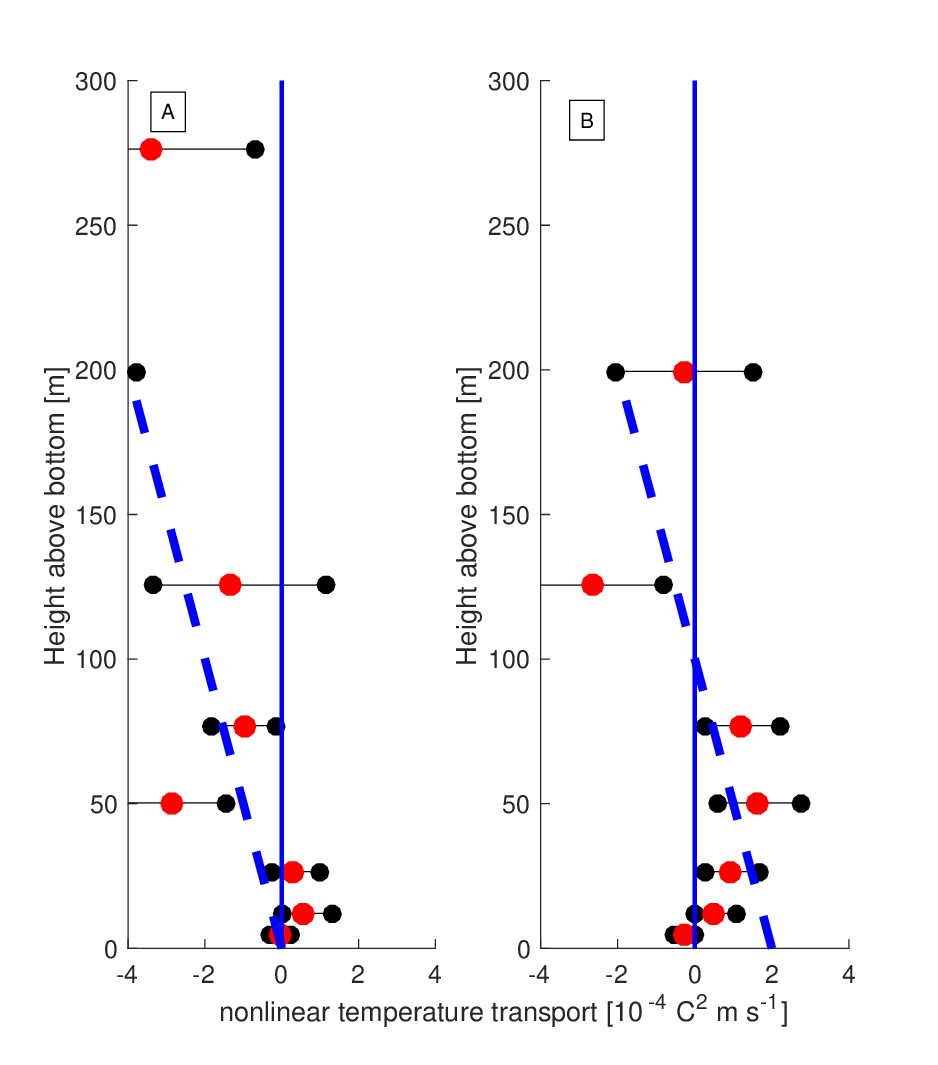}\\
\caption{ Triple correlations of temperature transport, $\overline{u_{\perp}\theta^{\prime 2}}$, associated with thalweg normal motions.  Red symbols are means and black 95\% confidence intervals based upon a bootstrap algorithm.  Panel A represents the first 2 fortnights, panel B the last 3.   The blue dashed line presents a gradient of $2\times10^{-6}$ C$^2$ s$^{-1}$ in height above coordinate normalized by the bottom slope.  }\label{fig:18}
\end{figure}

Our estimates of temperature flux have an interpretation in the context of a temperature evolution equation (\ref{eq:MeanReynolds}), in which the flux divergence can be interpreted as diapycnal motion (\ref{eq:upwelling}).  The temperature flux estimates also participate in a temperature variance equation:  
\begin{equation}
    (\partial_t + \overline{{\bf u}} {\bf \cdot \nabla}) \; \overline{ \theta^{\prime 2}} + {\bf \nabla} {\bf \cdot} \overline{{\bf u}^{\prime} \theta^{\prime 2}} + 2 \overline{{\bf u} \theta^{\prime}} {\bf \cdot} {\bf \nabla} \overline{\theta} = -\chi \; ,
\label{eq:temperaturevariance}
\end{equation}
in which $\theta$ is potential temperature and $\chi$ is the rate of dissipation of temperature variance.  The first term is the time rate of change of temperature variance following the mean flow, the second nonlinear transport terms, the third production and the right-hand side temperature variance dissipation by molecular processes.  Inertial subrange estimates of $\chi$ appear in figure \ref{fig:15}.  Note that $\chi$ is positive definite, $-\chi$ negative.  Our positive temperature fluxes represent a positive contribution to the left-hand-side (\ref{eq:temperaturevariance}) and are approximately an order of magnitude greater than $\chi$.  A crude assessment suggests an order of magnitude cancellation between anti-production and nonlinear transports are required.  Embedded in these nonlocal nonlinear transport terms is information about the boundaries.  We provide initial estimates of nonlinear transport terms in figures \ref{fig:17} and \ref{fig:18}.  

note to self:  compare with microstructure data and Alberto's chi's.  

\begin{figure}[t]
\noindent\includegraphics[width=0.48\textwidth]{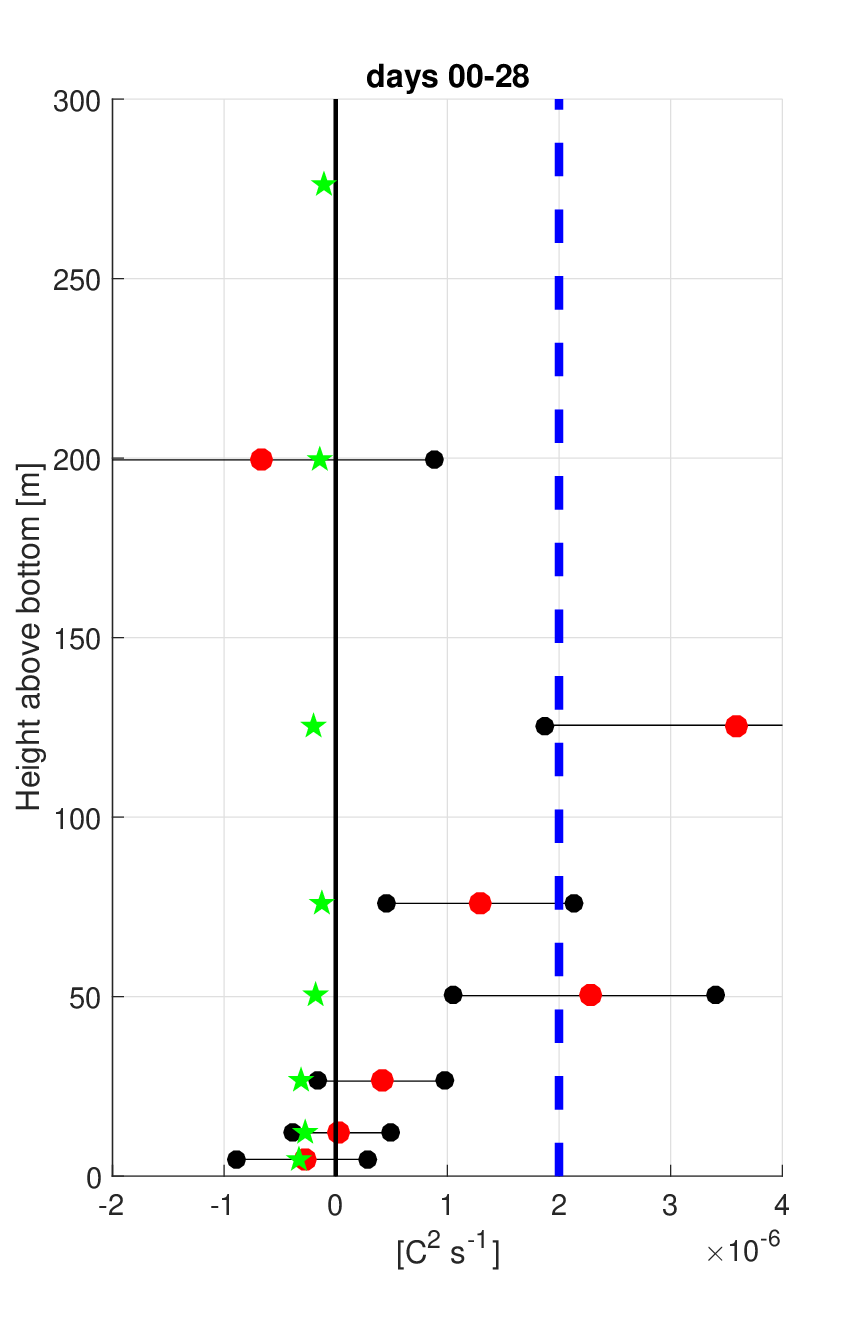}\\
\caption{ Vertical Production and Dissipation/2.  Production as red dots and dissipation as green pentagrams.  The blue dashed line provides a $2\times10^{-6}$ C$^2$ s$^{-1}$ carried over to figures \ref{fig:17} and \ref{fig:18}.  }\label{fig:16}
\end{figure}

\subsection{Step 4:  Off boundary evolution}

See Yuchen's work, comparison with Charlotte's work.  

Future work with MAVS.  

\section{Perspectives}\label{sec:perspectives}
\subsection{Looking to the Past}
The historical narrative concerning the failure of the 'one-dimensional model' is compelling.  Assumption 1 was invoked by \cite{phillips1970flows} in a laboratory study with mixing driven by an oscillating grid.  Here the turbulent fluxes are nearly isotropic and the boundary slope moderate, so that scaling the flux divergence to discard the horizontal components makes sense.  However, Assumption 1 is hardly the only way to express a no-normal flux boundary condition.  From a mathematical perspective, temperature fluxes projecting parallel to the boundary are sufficient.  This is far more sensible when one considers the oceanographic phenomenology and, indeed, finds observational support in figure \ref{fig:9}.   In reading the literature, one is reminded about the underlying Assumptions 1 and 2 through \cite{thorpe1987current}, but thereafter these caveats are not brought forward.  Thereafter, the model evolves to have the character of a fairy tale.  See \cite{polzin2022mixing}.

One can point to a similar intellectual juncture in \cite{garrett1979comment} and  \cite{armi1979reply}, cast as Armi v. Garrett in \cite{polzin2022mixing}.  If all one had was vertical profiles similar to the deep CTD cast in figure \ref{fig:3}, one can find sympathy for the perspectives enunciated in \cite{garrett1979comment}.  However, our results concerning upgradient fluxes bring forth a specific message about the intrinsic spatial non-locality of boundary mixing, and that appears to be far more in the corner of Armi's pioneering work concerning the interpretation of vertical profile data from the new-fangled Neil Brown Mark III CTD.  

\subsection{and Today}

The geometry of the canyon might be special, but our collective intuition is that the dynamics that give rise to up-gradient and non-local phenomena are not.  These are embedded in near-critical wave phenomena \citep{eriksen1982observations,slinn1996turbulent} and should appear in concert with bore-like phenomena on a planar sloping boundary \citep{van2006nonlinear, winters2015tidally}.  Similarly, flow separation is, nearly by definition, production of anomalies at the boundary and dissipation elsewhere.  

Our tools and the media for expressing our stories have assuredly changed over the course of 50 years.  However, it is not clear that our conceptual understanding has migrated far from the original story book.  

Earth system models do not have either the resolution or a sufficiently realistic representation of internal wave band motions to represent the no-normal flux bottom boundary condition as anything but one that brings isopycnals to project normal to the topography.  An example is provided by analysis the 1 km horizontal grid ROMS-CROCO based GIGATL simulation, \cite{huang2025submesoscale}.  In that model, despite having a 1 km horizontal resolution, the underlying topography is smoothed in a manner that completely eliminates critical slopes in the Brazil Basin, to the point that internal tide driven mixing is arguably underestimated by four orders of magnitude.  Yet the diapycnal mixing in the model is not highly unrealistic.  The anticipated internal wave mixing appears to be replaced by a submesoscale eddy process, and model graphics suggest the vorticity field is populated by point vortices characteristic of weakly damped idealized numerical simulations.  

There are likely key issues about PV forcing linked to ventilation of the near boundary region.  One wonders if the story book has simply been replaced by the Disney catalog.

\section{Conclusions}\label{sec:Conclusions}

We present covariance based estimates of temperature flux from a submarine canyon.  The direct estimates of the vertical temperature flux are up gradient, rather than down gradient, at 50-125 m m above canyon bottom for the first two fortnights following a dye release presented in \cite{wynne2024observations}.  We demonstrate that the vertical flux terms dominate the horizontal as concerns their projection across mean isotherms.  The vertical gradient of these vertical flux terms indicates a diathermal upwelling of $O(1 $ mm s$^{-1})$ over 50-200 m $H_{ab}$, similar to that documented by concentration weighted estimates of the migration of the dye across temperature isopleths.  

On timescales of a fortnight and longer, we find that the isotherms are weakly sloping to within 3 meters of the bottom, such that there no tendency for the isotherms to dip into the bottom boundary to satisfy a no-normal flux bottom boundary condition.  An expression of this is a strongly stratified near-bottom regime.  The three-dimensional temperature flux estimates indicate that the no-normal flux bottom boundary condition is satisfied by temperature fluxes being parallel to the boundary.  

Temperature-velocity cospectra indicate that semi-diurnal frequencies dominate the covariance.  Thus mixing and upwelling in this canyon results from a wave breaking process.  We argue that this wave breaking is a result of a near-resonant build-up of internal Kelvin waves \citep{ma2025standing} supported by the steeply supporting canyon sidewalls.  

The up-gradient vertical temperature fluxes are indicative of a spatially non-local process.  Initial estimates of a temperature variance budget indicate that the rate of dissipation of temperature variance, $\chi$, results from an order of magnitude subtractive cancellation between the up-gradient vertical temperature flux and the spatial divergence of nonlinear transport terms, i.e. triple correlations.  

These observations stand in direct contrast to the fundamental predictions for the spatial structure of the boundary layer originating with \cite{phillips1970flows} and \cite{wunsch1970oceanic}.  That model invokes vertical mixing only and is coupled to spatially {\em local} flux-gradient closures that bring isotherms normal to the topographic slope to satisfy the no-normal flux boundary condition.  The dipping of isotherms into the slope implies weakened stratification at the boundary.  We estimate that the scale height for this model is approximately 20 meters, far smaller than peak-to-peak semi-diurnal isotherm displacements in excess of 300 m, and significantly smaller than the locus of up-gradient temperature fluxes that require a spatially non-local closure.  Despite the prominence of this model in the literature \citep{polzin2022mixing}, we are not surprised by the failure.

\begin{figure}[t]
\noindent\includegraphics[width=0.48\textwidth]{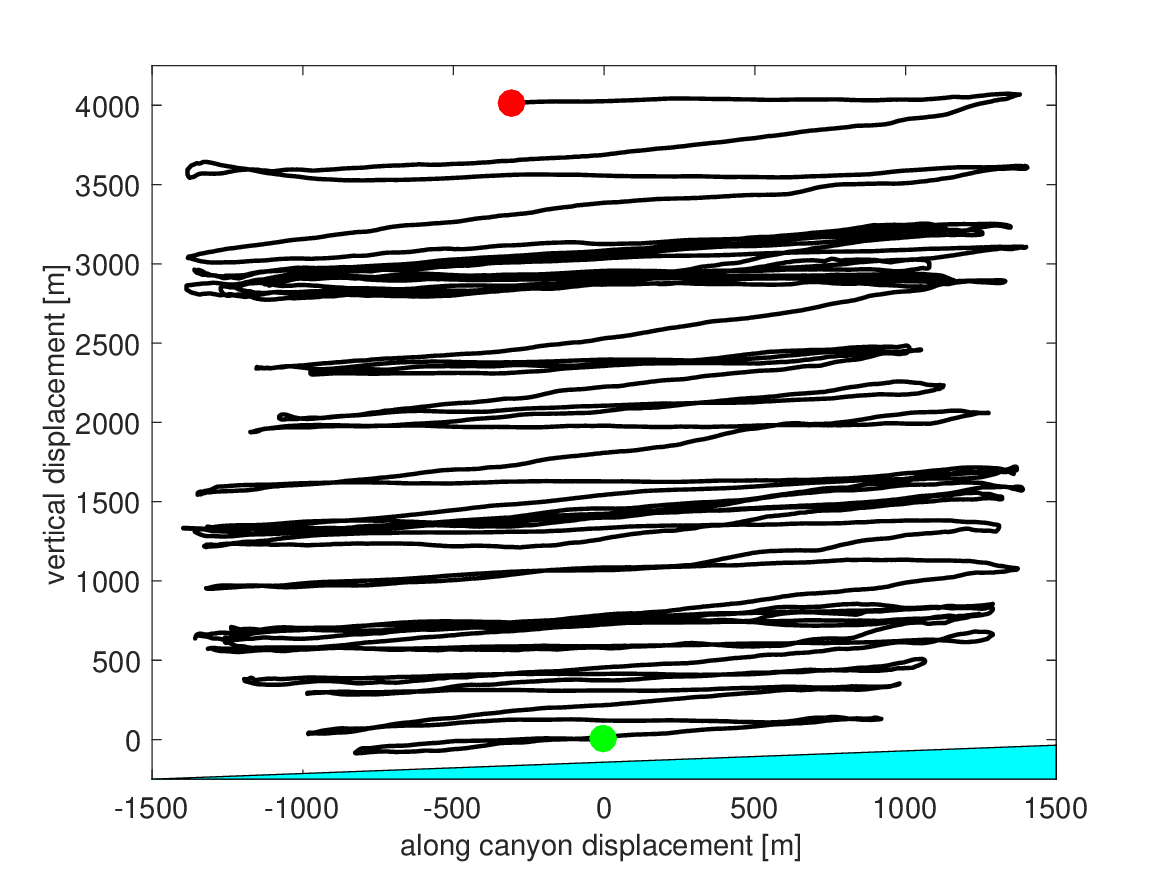}\\
\caption{Running integrals of band passed along-canyon and vertical velocities.  Unclear how to interpret this.  There is a vertical flux of horizontal momentum.  Maybe this is a GLM message.    }\label{fig:19}
\end{figure}

%

%

\clearpage
\acknowledgments
Kudos to the boat drivers and scientific support staff of the RRS Discovery.  
Financial support was provided by NSF grants OCE-  (KP); 

%
%
\datastatement

%






%



\bibliographystyle{ametsocV6}
\bibliography{BoundaryMixingNSF}

\end{document}